\documentclass[12pt]{article}
\usepackage{fullpage}
\usepackage{graphicx}
\newcommand{\be}{\begin{equation}}
\newcommand{\ee}{\end{equation}}

\newcommand{\bphi}{\mbox{\boldmath $\phi$}}

\font\mybb=msbm10 at 11pt

\def\bb#1{\hbox{\mybb#1}}

\def\Z {\bb{Z}}

\newcommand{\news}{\setcounter{equation}{0}}
\def\ben{\begin{equation}}
\def\een{\end{equation}}
\def\bea{\begin{eqnarray}}
\def\eea{\end{eqnarray}}

\begin{document}

\title{Vortex rings in ferromagnets \vskip 2cm }
\author{Paul Sutcliffe\\[10pt]
{\em \normalsize Department of Mathematical Sciences,
Durham University, Durham DH1 3LE, U.K.}\\
{\normalsize Email: p.m.sutcliffe@durham.ac.uk}}
\date{July 2007}
\maketitle

\begin{abstract}
Vortex ring solutions are presented for the Landau-Lifshitz equation, which
models the dynamics of a three-dimensional ferromagnet. The vortex rings
propagate at constant speed along their symmetry axis and are characterized
by the integer-valued Hopf charge. They are
stable to axial perturbations, but it is demonstrated that an easy axis
anisotropy results in an instability to perturbations which break the axial 
symmetry. The unstable mode
corresponds to a migration of spin flips around the vortex ring that leads
to a pinching instability and ultimately the collapse of the vortex ring. 
It is found that this instability does not exist for an isotropic ferromagnet.
Similarities between vortex rings
in ferromagnets and vortons in cosmology are noted.
\end{abstract}
\newpage

\section{Introduction}\news 

\ \quad 
In the continuum approximation the state of a ferromagnet is described
by a three-component unit vector $\bphi=(\phi_1,\phi_2,\phi_3)$ which
gives the local orientation of the magnetization. The dynamics of the
ferromagnet, in the absence of dissipation, 
is goverened by the Landau-Lifshitz equation
\be
\frac{\partial \bphi}{\partial t}=-\bphi \times 
\frac{\delta E}{\delta \bphi},
\label{ll}
\ee
where $E$ is the magnetic crystal energy of the ferromagnet and
units have been chosen in which the spin stiffness and
magnetic moment density of the ferromagnet are set to unity.

Work based upon a study of the conserved quantities
of the Landau-Lifshitz equation suggested \cite{Pap}
that vortex rings might exist which propagate at constant speed along their 
symmetry axis and are characterized by the integer-valued Hopf charge.
Analytical arguments have shown that the conservation of both 
spin flips and momentum is responsible for the existence of vortex rings,
and even solutions with zero Hopf charge exist \cite{Coo}.
Furthermore, numerical minimization methods, using a travelling wave
 ansatz with axial symmetry, 
have determined quantitatively the conditions
under which vortex rings exist, both for Hopf charge zero and one \cite{Coo}. 
However, it has yet to
be demonstrated that an initial condition which has the correct topological
properties will indeed evolve into a stable vortex ring in rigid motion along
its symmetry axis. Moreover, the stability properties to general non-axial
perturbations are unknown despite the fact that they are clearly of crucial 
importance if vortex rings are to be observed experimentally.

In this paper both these issues are addressed by performing numerical
simulations of the time-dependent three-dimensional Landau-Lifshitz equation,
both with and without an easy axis anisotropy.
It is shown that suitable initial conditions indeed yield axially symmetric
vortex rings, and examples with Hopf charges ranging from zero to five 
are presented.
The vortex rings are seen to propagate at constant speed along their 
symmetry axis and it is observed that the speed decreases as the Hopf charge
increases. However, it is found that 
 vortex rings in the anisotropic system are 
unstable to perturbations which break the axial symmetry. The unstable mode
corresponds to a migration of spin flips around the vortex ring that leads
to a pinching instability and ultimately the collapse of the vortex ring.
It is shown that this instability disappears in the isotropic limit.

The stability properties reported in this paper are clearly of importance
for any future attempts at an experimental 
observation of vortex rings, and reveal that anisotropy should be 
minimized in any experimental searches.

\section{Vortex ring dynamics}\news 

\ \quad
The system studied is that of a three-dimensional
ferromagnet with isotropic exchange interactions and an easy axis
anisotropy, in which case the energy is given by
\be
E= \frac{1}{2} \int (\partial_i\bphi\cdot\partial_i\bphi 
+ A (1-\phi_3^2)) \ d^3{\bf x},
\label{energy}
\ee
where $A\ge 0$ is the anisotropy parameter and the ground state
is $\bphi=(0,0,1)={\bf e}_3.$

With this energy the Landau-Lifshitz equation (\ref{ll}) becomes
\be
\frac{\partial \bphi}{\partial t}=\bphi \times 
(\partial_i\partial_i\bphi + A\phi_3{\bf e}_3).
\label{ll3}
\ee

In addition to the energy (\ref{energy}), the Landau-Lifshitz
 equation (\ref{ll3})
has two other conserved quantities which are relevant for the discussion
here, namely the number of spin flips,
 $N$, and the momentum ${\bf P}$, given by \cite{PT}
\bea
N&=&\int (1-\phi_3) \  d^3{\bf x},\label{spinflips}\\
P_i&=&\frac{1}{4}\epsilon_{ijk}\int x_j \epsilon_{klm}
\bphi\cdot\partial_l\bphi\times\partial_m\bphi \ d^3{\bf x}.
\label{mom}
\eea 

For the idealized case of a three-dimensional ferromagnetic
continuum without boundaries the finite energy condition is that
$\bphi\rightarrow {\bf e}_3$ as $|{\bf x}|\rightarrow \infty.$ 
This condition compactifies space to $S^3$,
so that at any given time the field is a map $\bphi: S^3 \mapsto
S^2.$ The relevant homotopy group is $\pi_3(S^2)=\Z$, 
so there is an associated integer conserved
topological charge $Q$, the Hopf charge,
which may be defined as a linking number, as follows.
The preimage of any point
on the target $S^2$ is generically a closed loop. Consider two such loops
corresponding to the preimages of two distinct points on the target $S^2,$ 
then these loops will be linked exactly $Q$ times for a field with Hopf
charge $Q.$

Twisted vortex rings provide examples of field configurations with
non-zero Hopf charge, and will be described shortly. 
First of all, it is helpful
to recall the properties of stationary topological soliton solutions of the
Landau-Lifshitz equation in two spatial dimensions \cite{PZ}.
In two dimensions the finite energy condition compactifies the plane to $S^2$
so that at any given time the field is a map $\bphi: S^2 \mapsto
S^2.$ Now the relevant homotopy group is $\pi_2(S^2)=\Z$, 
so there is again an associated integer conserved charge $q,$ which
is simply the winding number or degree of the map $\bphi$
and is given by
\be
q=\frac{1}{4\pi}\int \bphi\cdot(\partial_1\bphi\times
\partial_2\bphi)\ d^2{\bf x}.
\label{lumpch}
\ee

Let $r$ and $\theta$ be polar coordinates in the plane, then the
soliton located at the origin with positive charge $q$ has the form
\be
\phi_1+i\phi_2=e^{i(q\theta-\omega t+\alpha)}\sin g, \quad 
\phi_3=\cos g,\label{ansatz2}
\ee
where the constant $\alpha$ is an internal phase and 
$g(r)$ is a real radial profile function which satisfies the 
ordinary differential equation
\be
g''+\frac{1}{r}g'-(A+\frac{q^2}{r^2})\sin g\cos g +\omega \sin g =0
\label{profile}
\ee
together with the boundary
conditions $g(0)=\pi$ and $g(\infty)=0.$ 
The location of the 
soliton is the origin since here  the field takes
the value $\bphi=-{\bf e}_3,$ which is antipodal to the
vacuum value. 
Note that this solution is stationary not static, as the
first two components of the field rotate with a constant positive 
angular frequency $\omega.$ 
Although the field depends on the angular coordinate $\theta$
it is axially symmetric in the sense that a spatial rotation can be
compensated by the action of the global $O(2)$ symmetry of the theory,
which rotates the first two components of $\bphi.$ 

A linearization of the profile function equation
(\ref{profile}) reveals that a solution satisfying the required boundary
conditions certainly must obey the constraint $\omega\le A.$ In fact 
(for each $\alpha$) there
is a one-parameter family of solutions labelled by $\omega,$ or alternatively
by the number of spin flips $N,$ with a nonlinear
 relationship between these two quantities with the property that as $\omega$
decreases then $N$ increases. 

For an isotropic ferromagnet 
($A=0$) the situation is somewhat different since in this case 
the constraint requires
that $\omega=0,$ so the solutions are static. However, there is still 
a one-parameter family of solutions labelled by $N$ and this follows from
the scaling symmetry $r\mapsto \lambda r$ of the profile function equation
(\ref{profile}) present when $A=\omega=0.$ In this isotropic case the 
profile function equation can be solved in closed form and the resulting
static solutions are the well-known magnetic bubbles.
In fact, in the isotropic case the static Landau-Lifshitz equation is
identical to the static equation of the $O(3)$ sigma model and in the plane
all multi-soliton solutions, not just those with axial symmetry, 
can be found explicitly in closed form using
rational functions \cite{BP}.

As described later, vortex rings have a constant rotation of the
first two components of $\bphi.$ They are therefore easiest to 
understand in
the anisotropic case, since they are essentially constructed from similar
stationary two-dimensional solitons. As discussed above, 
in the isotropic limit there are only static, not stationary, two-dimensional
solitons, and this complicates the situation. For simplicity, the anisotropic
system will be investigated initially, though later the isotropic system
will be studied too, since it turns out that there is an important difference
between the stability of vortex rings in these two cases.   
The detailed computations of Ref.\cite{Coo} were mainly performed
with $A=0$ but it was noted that vortex rings are apparently more favourable
with $A>0,$ in the sense that they exist for a larger range of parameters.
In the first part of the present  
study the anisotropy will be included by setting $A=1,$ which can be chosen
without loss of generality by a rescaling of the length and time units.  

Given a two-dimensional stationary soliton with charge $q$ this can be
embedded into the three-dimensional theory along a circle to produce
a vortex ring. If the internal phase $\alpha$ is rotated through an angle
$2\pi m$ as the soliton travels around the circle once then the resulting
vortex ring has Hopf charge $Q=qm.$  Only the simplest 
case of a single $q=1$ soliton will be investigated here, thus $Q=m.$    

Explicitly, using cylindrical polar coordinates $\rho,\chi,z$ 
the initial field configuration has the form
\be
\phi_1+i\phi_2=e^{iQ\chi}\frac{(\rho-\rho_0+i(z-z_0))}{R}\sin g,\quad
\phi_3=\cos g,
\label{start}
\ee
where $R=\sqrt{(\rho-\rho_0)^2+(z-z_0)^2},$ and $g(R)$ is a 
monotonically decreasing
profile function with the properties that $g(0)=\pi$ and $g(R)=0$
for $R\ge \rho_0.$ The parameter $\rho_0$ is the radius of
the circle defining the vortex ring location and $z_0$ is the position
of this circle in the spatial direction orthogonal to the plane
of the circle (which is taken to be the third spatial direction $z=x_3$).
The boundary condition $g(R)=\pi$ ensures that
on the circle $\rho=\rho_0, \ z=z_0$ the field takes the antipodal
vacuum value $\bphi=-{\bf e}_3,$ and the condition that
$g(R)=0$ for $R\ge \rho_0$ makes the field (\ref{start}) 
well-defined inside this circle (which in particular includes the origin).

By symmetry arguments it can be seen that such an initial 
configuration has $P_1=P_2=0$ and $P_3>0,$ so the momentum points 
along the symmetry axis of the vortex ring. This observation was
first made in Ref.\cite{Pap} and motivated the suggestion that vortex
rings might exist that propagate along their symmetry axis. The expectation
is that the conserved momentum provides a dynamical stability that prevents
the radius of the vortex ring from shrinking. The constrained energy 
minimization computations of Ref.\cite{Coo} confirm this expectation and
reveal that the conservation of both momentum and spin flips is enough
to provide a dynamical stability for both the radius and the thickness
of the vortex ring. It was found that vortex rings exist providing
$P_3/N^{2/3}$ is sufficiently large. Essentially this is the requirement
that the radius of the vortex ring is above a critical value 
determined by its thickness. If this combination is too small then the
radius of the ring shrinks to zero.

To construct a suitable function $g(R)$ for use in the ansatz (\ref{start}) 
one approach would be to choose
a rotation frequency $0<\omega<1$ (which determines the size of the embedded
two-dimensional soliton and hence the thickness of the vortex ring) and
solve the profile function equation (\ref{profile}) with $R$ replacing $r$
and the boundary condition $g(r=\infty)=0$ replaced by $g(R=\rho_0)=0.$
Providing the radius of the vortex ring $\rho_0$ is significantly larger than
the size of the embedded soliton then this approach ensures that the embedded
soliton is similar to the true two-dimensional soliton. However, given that
one of the objectives of the present study is to investigate whether vortex
rings can emerge from a range of initial conditions, and to test their
stability properties, it is useful to begin with a profile function that is
only a reasonable approximation to a two-dimensional soliton. 
 
\begin{figure}[t]
\begin{center}
\leavevmode
\includegraphics[width=16cm]{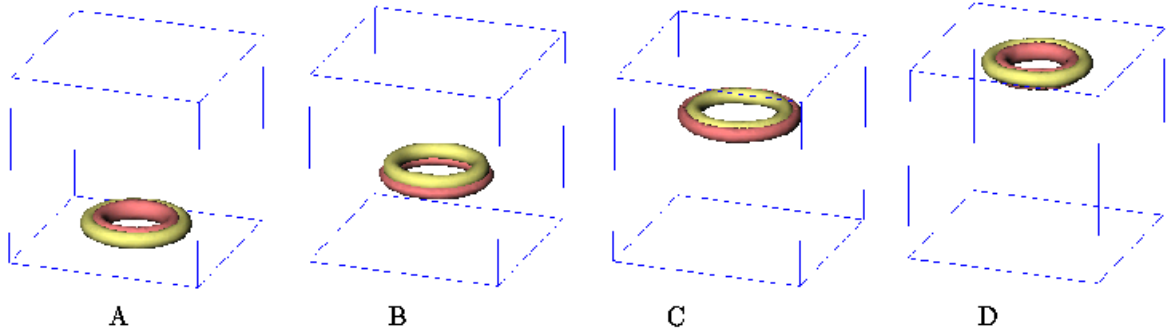} 
\caption{Tubular surfaces indicating the preimages of the points
$\bphi=(0,0,-1)$ (light surface) and $\bphi=(-1,0,0)$ (dark surface), at times
$t=0, 39,  78,  109, $ for a vortex ring with Hopf charge $Q=0.$
}
\label{fig-Q0}
\vskip 0cm
\end{center}
\end{figure}

\begin{figure}[t]
\begin{center}
\leavevmode
\includegraphics[width=16cm]{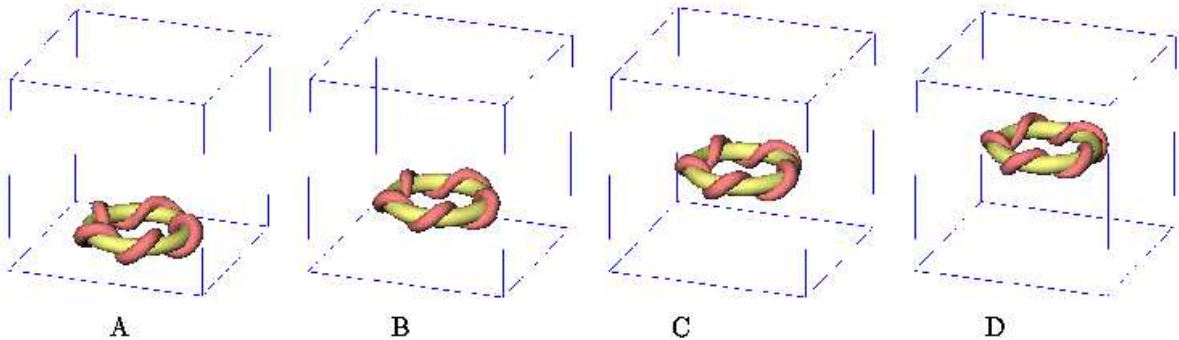} 
\caption{As Figure \ref{fig-Q0} but with Hopf charge $Q=5.$
}
\label{fig-Q5}
\vskip 0cm
\end{center}
\end{figure}

A range of profile functions and vortex ring radii have been investigated, 
and all produce
qualitatively similar results. The results presented in Figure \ref{fig-Q0}
(for $Q=0$) and Figure \ref{fig-Q5} (for $Q=5$) 
are for a vortex ring with radius $\rho_0=6$ 
and profile function
\be
g(R)=\pi\left(1-\frac{R}{5}\right)^3 \Theta(5-R),
\label{approx1}
\ee
where $\Theta$ is the Heaviside step function. This profile function is 
a reasonable approximation to the exact solution of the profile
function equation (\ref{profile}) with a frequency $\omega=0.56.$
For this initial condition the spin flips and momentum are $N=407$
and $P_3=1403.$

The Landau-Lifshitz equation (\ref{ll3}) was solved numerically using an
explicit finite difference scheme which is fourth-order accurate in the
spatial derivatives and second-order accurate in the time derivative.
The grids used contain $101^3$ points with a lattice spacing 
$\Delta x=0.25$ and a time step $\Delta t=0.003125,$ though larger grids
with $151^3$ points were sometimes used, as were smaller lattice spacings,
to test the numerical accuracy. Periodic boundary conditions were applied
in all the simulations.

\begin{figure}[t]
\begin{center}
\leavevmode
\includegraphics[width=10cm]{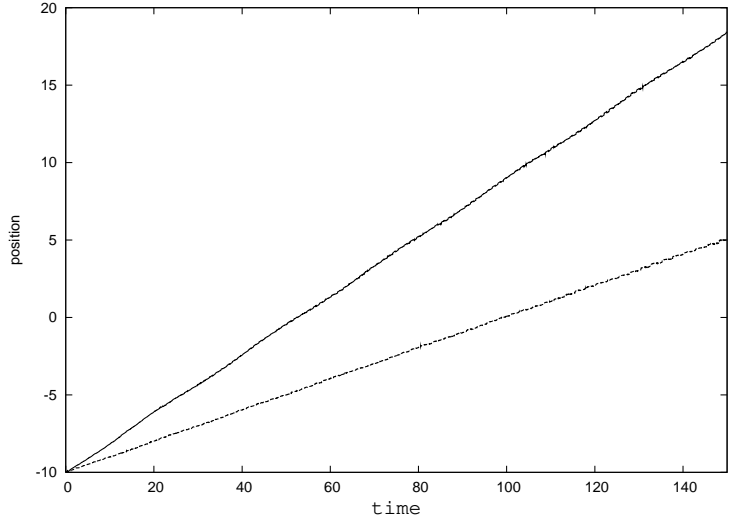} 
\caption{The position of the vortex ring as a function of time, for
Hopf charge $Q=0$ (upper curve) and $Q=5$ (lower curve).
}
\label{fig-speed}
\vskip 0cm
\end{center}
\end{figure}

In Figure \ref{fig-Q0} the results of a simulation are presented where 
 the
initial conditions are as described above with a 
vortex ring position $z_0=-10.$
The results are presented by displaying the location of the vortex ring,
which is the curve given by the preimage of the point $\bphi=(0,0,-1).$
To aid visibility this curve is thickened out to a tubular surface 
(light surface) in  Figure \ref{fig-Q0} by plotting the isosurface $\phi_3=0.$
The Hopf charge can be seen by considering the linking number of this
curve with the preimage of any other point, which has been chosen to be
$\bphi=(-1,0,0),$ and again this is thickened out to a tubular surface 
(dark surface) in  Figure \ref{fig-Q0} by plotting the isosurface $\phi_1=-0.5.$

In Figure \ref{fig-Q0} the tubular surfaces are displayed at  times
$t=0, 39, 78,  109. $ A glance at  Figure \ref{fig-Q0}A reveals that 
initially the
dark tube is similar to the light tube but has a smaller radius, so the
linking number is indeed $Q=0.$ The evolution displayed at later times
confirms that the whole configuration drifts upwards, whilst the dark tube
rolls around the light tube from the inside to the outside in a periodic
motion. This simply reflects the stationary properties of the two-dimensional
soliton, where the first two components of the field rotate with an
angular  frequency
$\omega.$ As mentioned above, the profile function given above can be
reasonably compared to that of the two-dimensional soliton with $\omega=0.56$
so the expected period of the rolling tube motion is roughly $T\approx 11,$
which is in good agreement with the results of the simulation 
presented in Figure \ref{fig-Q0}. 

In Figure \ref{fig-speed} the position of the $Q=0$ vortex ring 
is shown (upper curve) as a function of time. 
Note that this is the position in the $x_3$ direction, 
which is along the symmetry axis, and the periodicity of the 
simulation domain has been taken into account when computing this
position. This plot demonstrates that, to a very good approximation,
the vortex ring indeed travels with a constant velocity. 
This velocity is roughly just less than $v\approx 0.2,$ 
so in the time taken for the
vortex ring to move a distance equal to its diameter the two-dimensional
soliton forming the vortex ring has made over five revolutions. Thus the
translational motion of the vortex ring is somewhat slower than the 
rotational motion. A careful inspection of the upper curve in
Figure \ref{fig-speed} reveals that there are very small amplitude wiggles
superimposed onto a linear motion. These wiggles correspond to the very slight
oscillation in the radius of the vortex ring that results from the fact
that the initial conditions are only constructed approximately, and provides
evidence for the stability of the vortex ring to axially symmetric 
perturbations. The simulations presented above were run for a much longer length of time than
shown here, so that the vortex ring crosses the periodic boundary of the
simulation grid several times, and it continues to propagate at constant speed
in the same manner.

Figure \ref{fig-Q5} displays the results of a simulation identical to the
previous one discussed above, except that the initial condition has
$Q=5$ rather than $Q=0.$ It is clear to see that the Hopf charge is $Q=5$
because the dark tube winds around the light tube five times.
 This vortex ring behaves very much like the
$Q=0$ vortex ring, except that the propagation speed has been roughly halved to
$v\approx 0.1,$ as can be verified from  Figure \ref{fig-speed} 
where the lower curve displays the position against time. 
Performing a number of simulations with different values of the Hopf charge 
confirms that it is a general feature that increasing only the Hopf charge, 
while leaving
all other parameters unchanged, leads to a reduction
in the propagation speed. 
 Note that for a vortex ring with $Q\ne0$ the internal 
rotation of the two-dimensional soliton forming the vortex ring 
is equivalent to a spatial rotation of the whole vortex ring around its
symmetry axis, so either interpretation is equally valid, but for
$Q=0$ the interpretation as a rotation of the ring around its
symmetry axis is not allowed.

\begin{figure}[t]
\begin{center}
\leavevmode
\includegraphics[width=16cm]{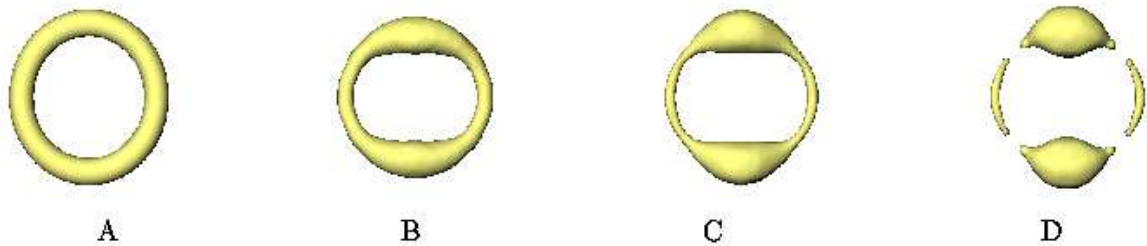} 
\caption{The isosurface $\phi_3=0$ at times 
$t=0, \ 37.5, \ 50, \ 55.625, \ $ for a $Q=0$ vortex ring with
an initial $10\%$ squashing in the $x_1$ direction.
}
\label{fig-P2}
\vskip 0cm
\end{center}
\end{figure}

The results presented so far, together with those of Ref.\cite{Coo}, suggest
that vortex rings (with sufficient momentum) are stable to axially symmetric
perturbations. To investigate the stability properties to general non-axial
deformations a rescaling perturbation is applied to the initial conditions 
to squash the vortex ring by $10\%$ in the $x_1$ direction (recall that
the symmetry axis is the $x_3$ direction). The simulation presented
in  Figure \ref{fig-P2} is on a slightly larger grid containing $151^3$ points,
but again with $\Delta x=0.25.$ The vortex ring has zero Hopf charge  
and is a little larger and thicker than in the previous simulations, with
the number of spin flips and momentum being $N=2186$
and $P_3=5163.$ However, a number of simulations
have been performed with various grid sizes and lattice spacings, using 
vortex rings of varying diameters, thickness and Hopf charge, and all the
results are qualitatively similar. The isosurface $\phi_3=0,$ which
is shown at times $t=0,37.5,50,55.625,  $ in Figure \ref{fig-P2}
illustrates not only the location of the vortex ring but also its
thickness, which corresponds to a measure of the local spin flip density 
in the two-dimensional soliton forming the vortex ring. 
Clearly there is an instability associated with
 a migration of spin flips around the vortex ring that leads
to a pinching of the vortex ring at certain locations and a compensating 
fattening at other locations in order to conserve
the total number of spin flips. In the simulation presented in  
Figure \ref{fig-P2} the initial squashing perturbation preserves a $C_2$
symmetry so the spin flips accumulate at two symmetrical points.

The pinching instability can also be observed in a slightly simpler
setting, namely in a straight vortex string, rather than a vortex ring. 
This provides a reasonable approximation to a segment of the vortex  
ring if the radius is much larger than the thickness. Numerical simulations
on a range of vortex strings with varying thickness and windings, 
have revealed that
all suffer from a pinching instability, and that thicker vortex strings
take longer to pinch, but no evidence was found that for a sufficient
thickness the instability is removed. For straight vortex strings it would
be interesting to confirm this numerically by a study of the negative mode
 as a function of the string thickness and winding,
using a linearized eigenvalue computation.
However, it seems unlikely that the instability is lost at a critical
thickness, since the unstable mode essentially leads to the formation
of magnetic solitons, which in the stationary case are spherically
symmetric \cite{KIK}. It is simple geometry, based on concepts of volume and
surface area, that suggests a preference for the spin flips to form
lumps rather than tubes. 

It is an important observation that the unstable mode associated with the 
migration of spin flips proceeds towards the formation of magnetic solitons.
This is crucial because stationary magnetic solitons exist only for an 
anisotropic ferromagnet and not in the isotropic limit, as discussed
earlier for two-dimensional topological solitons. This suggests that the
decay mode might be removed for vortex rings in an isotropic ferromagnet.
Although the instability has been discovered using dynamical simulations,
it turns out that it can also be studied using a constrained energy
minimization method, and this is a slightly better approach since
long lived dynamical oscillations do not have to be addressed.
This approach is described in the following section, where it is used to 
investigate the stability of vortex rings in both the anisotropic and isotropic
case.

\section{Constrained energy minimization}\news 

\ \quad
Axially symmetric vortex rings were constructed in Ref.\cite{Coo}
using a constrained energy minimization method. In this section 
a similar approach is applied but the previous restriction to axially 
symmetric fields is removed, which enables the investigation of any
possible unstable modes. 
 
There is an ansatz which is consistent with
the equations of motion and describes a solution that propagates at
constant velocity ${\bf v},$ while the first two components of $\bphi$  
rotate with a
frequency $\omega.$ Explicitly, the ansatz reads
\be
\phi_1+i\phi_2=(\widetilde\phi_1({\bf x}-{\bf v}t)
+i\widetilde\phi_2({\bf x}-{\bf v}t))e^{-i\omega t}, \ \
\phi_3=\widetilde\phi_3({\bf x}-{\bf v}t).
\label{trav}
\ee
Substituting this ansatz into the equation of motion
(\ref{ll3}) leads to a partial differential equation for
$\mbox{\boldmath $\widetilde\phi$}=
(\widetilde\phi_1,\widetilde\phi_2,\widetilde\phi_3).$
This partial differential equation
has a variational formulation, which is useful in computing its
solutions. Let $\widetilde E({\bf P},N)$ be the minimal value of the 
energy $E,$ given by (\ref{energy}), for fixed
values of the momentum ${\bf P}$ and number of spin flips $N$.
Then the solution of this constrained minimization problem is
precisely the function 
$\mbox{\boldmath $\widetilde\phi$}$
corresponding to the values
\cite{TW}
\be
\omega=\frac{\partial\widetilde E}{\partial N}\bigg\vert_{\bf P}, \ \
v_i=\frac{\partial\widetilde E}{\partial P_i}\bigg\vert_N.
\label{speed}
\ee

Vortex ring solutions are computed by numerically solving
this constrained energy minimization problem using a simulated annealing 
algorithm with the constraints
on $N$ and ${\bf P}$ imposed using a penalty function method.
As with the earlier dynamical simulations, the momentum is chosen
to point along the $x_3$ axis. The energy is
discretized using a finite difference method and 
the grid has the same dimensions as in the dynamical simulations,
that is, it contains $151^3$ points and has a lattice spacing
$\Delta x=0.25.$

 \begin{figure}[t]
\begin{center}
\leavevmode
\includegraphics[width=16cm]{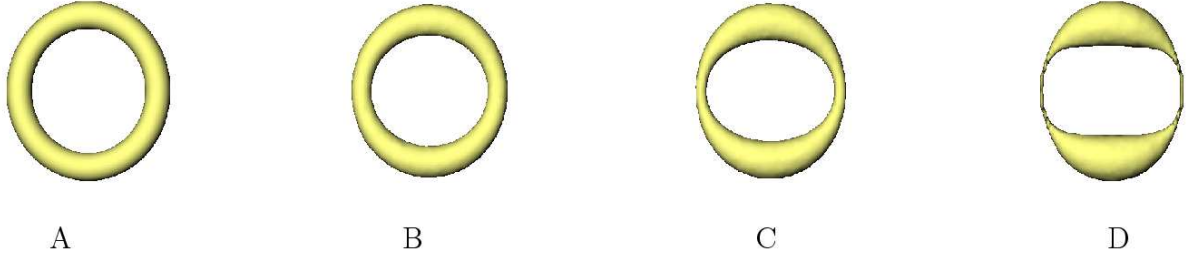} 
\caption{The isosurface $\phi_3=0$ during increasing stages of a 
 simulated annealing relaxation
of a $Q=1$ vortex ring in the anisotropic system with $A=1.$ 
The spin flips and momentum are constrained to be $N=2186$ and $P_3=5163.$
The initial condition is squashed by $10\%$ in the $x_1$ direction and
leads to a pinching instability.
}
\label{fig-saA}
\vskip 0cm
\end{center}
\end{figure}

 \begin{figure}[t]
\begin{center}
\leavevmode
\includegraphics[width=7.5cm]{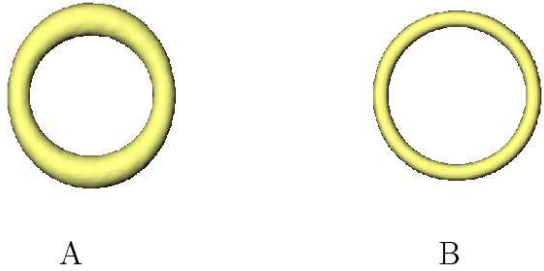} 
\caption{
The isosurface $\phi_3=0$ during increasing stages of a 
 simulated annealing relaxation
of a $Q=1$ vortex ring in the isotropic system ie. $A=0.$ 
The spin flips and momentum are constrained to be $N=2186$ and $P_3=5163.$
The initial condition is a slightly pinched and squashed vortex ring
but leads to an axially symmetric stable solution.
}
\label{fig-saB}
\vskip 0cm
\end{center}
\end{figure}

Initial conditions can again be created using the axial ansatz
(\ref{start}), with an optional squashing in the $x_1$ direction
to break the axial symmetry. Figure \ref{fig-saA} displays the results
at increasing stages of a simulated annealing energy relaxation computation,
for the system with anisotropy parameter $A=1.$
The initial condition 
(Figure \ref{fig-saA}A)
is identical to that used in the dynamical simulation
presented in Figure \ref{fig-P2} except that now $Q=1$ rather than $Q=0.$
The pinching instability is clearly displayed in Figure \ref{fig-saA}
under constrained energy minimization, and the qualitative features 
are very similar to those found in the dynamical simulations (compare
Figure \ref{fig-P2}).

Performing similar annealing simulations in the isotropic model ($A=0$)
does not produce a pinching instability. As an example, consider 
the results presented in Figure \ref{fig-saB}. The initial condition 
 (Figure \ref{fig-saB}A) is
taken from an early stage of the annealing simulation with $A=1$ displayed in 
 Figure \ref{fig-saA}, and therefore initially has a slight
pinching in addition to the squashing. However, with $A=0$ 
the pinching is reversed and the annealing process
results in the stable energy minimizing axially symmetric vortex ring displayed
in  Figure \ref{fig-saB}B. There is clearly no pinching instability
in the isotropic system. Note that the thickness of the axial vortex ring 
with $A=0$ appears less than in the $A=1$ system, because of the different
localization profiles of solitons in the two systems.

The vortex ring solution of Figure \ref{fig-saB}B, generated by energy 
minimization, has been used as an initial condition in a dynamical
simulation of the type discussed in section 2, but now in the isotropic
system ($A=0$). The results confirm
that the vortex ring propagates with an unchanged form at constant
speed, in a similar manner to the $A=1$ results. Perturbed vortex rings
with broken axial symmetry have also been used as initial conditions in the
dynamical simulations and the results confirm the energy minimization 
computations, in that no pinching instabilities are seen.

\section{Conclusion}\news 

\ \quad
Propagating vortex ring solutions of the Landau-Lifshitz equation have been
presented and their stability properties investigated.
It has been shown that for an anisotropic ferromagnet there
is a pinching instability which leads to the collapse of the vortex ring,
but that this instability disappears for an isotropic ferromagnet.
It is possible that even in the anisotropic case there could be
parameter values $N,P_3,Q$ for which vortex rings do not suffer from the 
pinching instability. By scaling symmetry, the properties of vortex rings
for a given $Q$
depend upon the parameter combinations $p=P_3/N^{2/3}$ and $a=AN^{2/3}.$
The computations performed in the anisotropic system have been
limited to reasonably large values of both $p$ and $a$
(the simulation presented in Figure \ref{fig-saA} corresponds to 
$p\approx 30.7$ and $a\approx 168$).
A comprehensive
search of parameter space is required to determine if there are any 
parameter regimes
with small $a\ne 0$ that allow stable rings, 
but this is beyond the scope of the 
present study.
As mentioned earlier, when the anisotropy parameter $A$ is non-zero it can 
be set to unity, without loss of generality, by rescaling length and time 
units. However, this means that an investigation of a wide parameter 
range of $a$ requires substantial variations in $N$ and this is
problematic numerically, because a variety of size scales must be accomodated. 
The alternative is to keep $N$ fixed and vary $A.$ For example, the
simulation presented in Figure \ref{fig-saA} has been repeated but with $A=0.1$
rather than $A=1$, corresponding to a less anisotropic regime
with $a\approx 16.8$. The pinching instabilty is again revealed, but on
a time scale that is an order of magnitude larger than in the previous case.
It is therefore difficult to determine whether the instability is removed for
a small but non-zero value of $a,$ or only in the limit $a=0.$
In either case, if the anisotropy is small, then any potential
pinching instability will only develop on large timescales
and therefore may not be important, since the lifetime of any excitation will
be limited by dissipation, which has been neglected here.
These results suggests that any future attempts at an experimental 
observation of vortex rings will most likely succeed in ferromagnets
where the anisotropy is as weak as possible.

In a different system to the one studied here (though it is
related and has the same field content) stable static Hopf solitons exist which
form knotted configurations \cite{BS5,Su7}. 
It would be very interesting if similar propagating knotted Hopf solitons
existed in
ferromagnets, thereby providing a physical system in which knots may be 
observed experimentally. Clearly the anisotropic system is unlikely to
support such solitons due to the pinching instability, but there is a 
possibility that the isotropic system might. A variety of knotted and
linked vortex string 
field configurations with various Hopf charges can be generated
using an ansatz involving rational maps \cite{Su7}. Such fields have been
 used as initial conditions in the constrained energy minimization method 
discussed
in the previous section, but unfortunately this has, as yet, not produced
any stable knotted or linked solutions. The relevant obstacle appears
to be the fact that to minimize energy with a constrained momentum
it is favourable to localize the field close to a plane which is perpendicular
to the momentum; as in vortex rings. As a result, initially knotted
fields relax by flattening and this leads to vortex string reconnections and
ultimately the formation of vortex rings. 

In addition to the conserved quantities of spin flips $N$ and momentum 
${\bf P}$ there is also a conserved angular momentum ${\bf L}$ \cite{PT}.
For axially symmetric vortex rings the angular momentum points along
the symmetry axis and has the value $L_3=QN,$ which reflects the fact
mentioned earlier that for a  vortex ring with $Q\ne 0$
 the internal 
rotation of the two-dimensional soliton forming the vortex ring 
is equivalent to a spatial rotation of the whole vortex ring around its
symmetry axis. To prevent knotted initial configurations from relaxing
to axially symmetric vortex rings a constraint on $L_3$ can be imposed
with $L_3\ne QN.$ However, the result of such relaxations simply 
tends to produce vortex rings which are elliptical rather than circular.
In summary, the current evidence suggests that stable propagating
knotted vortex strings appear unlikely to exist, though 
it remains an open problem as there are a large
range of parameter values for $N,{\bf P},{\bf L},Q$ and only a handful
of specific examples have currently been investigated. 
 
Finally, it is interesting to note that there is a comparison to be made
between vortex rings in ferromagnets and similar objects
in relativistic field theories. Non-topological straight string solitons 
carrying Noether charge suffer from a similar pinching instability
\cite{CKL}, with a migration of charge along the string 
leading to the conversion of the string into non-topological solitons.
Vortons, which arise in cosmological applications, are closed loops
of superconducting cosmic strings carrying both current and charge. 
They have a number of similarities with the vortex rings studied here, and   
in particular there are parameter regimes in which a related 
pinching instability exists \cite{LS}. Thus vortex rings in ferromagnets
may be yet another example in which analogues of cosmological phenomena
may be studied in condensed matter systems in the laboratory. 

\section*{Acknowledgements}\news
 
Many thanks to Nigel Cooper and Theodora Ioannidou for useful discussions.
This work was supported by the PPARC special programme
grant ``Classical Lattice Field Theory''.
The parallel computations were performed on COSMOS at the National
Cosmology Supercomputing Centre in Cambridge.

\end{document}